\documentclass{aastex}
\usepackage{graphicx}

\shorttitle{Extrasolar Binary Planets II}
\shortauthors{Ochiai, Nagasawa, Lewis \& Ida}

\newcommand{\bun}[2]{\left(\frac{#1}{#2}\right)}

\begin{document}

\title{Extrasolar Binary Planets II:\\
Detectability by Transit Observations}

\author{K. M. Lewis$^1$, H. Ochiai$^2$, M. Nagasawa$^3$,  and S. Ida$^1$}
\affil{1) Earth-Life Science Institute, Tokyo Institute of Technology, 2-12-1 Ookayama, \\
Meguro-ku, Tokyo 152-8550, Japan}
\affil{2) Earth and Planetary Sciences, Tokyo Institute of Technology, 2-12-1 Ookayama, \\
Meguro-ku, Tokyo 152-8551, Japan}
\affil{3) Interactive Research Center of Science,
Tokyo Institute of Technology, 2-12-1, Ookayama, \\
Meguro-ku, Tokyo 152-8551, Japan
}

\email{nagasawa.m.ad@m.titech.ac.jp}

\begin{abstract}
We discuss the detectability of gravitationally bounded pairs of  gas-giant planets (which we call ``binary planets") in extrasolar planetary systems that are formed through orbital instability followed by planet-planet dynamical tides during their close encounters, based on the results of N-body simulations by Ochiai, Nagasawa and Ida (Paper I).  Paper I showed that the formation probability of a binary is as much as $\sim 10\%$ for three giant planet systems that undergo orbital instability, and after post-capture long-term tidal evolution, the typical binary separation is 3--5 times the sum of physical radii of the planets.  The binary planets are stable during main sequence lifetime of solar-type stars, if the stellarcentric semimajor axis of the binary is larger than 0.3 AU.    We show that detecting modulations of transit light curves is the most promising observational method to detect binary planets.  Since the likely binary separations are comparable to the stellar diameter, the shape of the transit light curve is different from transit to transit, depending on the phase of the binary's orbit.  The transit durations and depth for binary planet transits are generally longer and deeper than those for the single planet case.  We point out that binary planets could exist among the known inflated gas giant planets or objects classified as false positive detections at orbital radii $\ga 0.3$ AU, propose a binary planet explanation for the CoRoT candidate SRc01 E2 1066, and show that binary planets are likely to be present in, and could be detected using Kepler-quality data.
\end{abstract}



\section{INTRODUCTION}
More than 7000 transit-like light curves have been obtained by Kepler observations\footnote{http://exoplanetarchive.ipac.caltech.edu/cgi-bin/ExoTables/nph-exotbls?dataset=cumulative}.  Among them, more than 3000 objects are identified as planetary candidates and more than 2000 objects are false positives.  Over 800 of them show transit depth comparable to that caused by gas giant planets.  In this paper, we point out that some of the known inflated gas giant planets and objects classified as false positive detections could be gravitationally bounded pairs of gas-giant planets, which we call ``binary planets."

Orbital stability of exomoons around gas-giant planets have been studied, and these moons are stable except for those around close-in gas giants \citep[e.g.,][]{Namouni,gong}.  ``Habitability" of the moons around gas giants in habitable zones has also been discussed \citep[e.g.,][]{Williams, Heller12, Heller13}.  In addition, the observational detectability of exomoons has been discussed
and many detection methods have been proposed \citep[e.g.,][]{SS99,SSS,kippingttv,kippingb,SA09,Kaltenegger,kipping12,zgy,Lewis13}.  Gas giants that are distant from their host star may commonly have moons, because regular moons are formed in circumplanetary disks \citep[e.g.,][]{Mosqueira03a,Mosqueira03b,Canup06,Sasaki10} and the formation of these disks is a part of gas-giant planet formation.  However, since \citet{Canup06} proposed that the maximum mass of a moon may be $\la 10^{-4} M_{\rm p}$ where $M_{\rm p}$ is the planet mass, detection of these moons is not easy.

On the other hand, it is relatively easy to distinguish binary gas giants from single gas giants,
because the companions are as large as the primary.  \citet[][hereafter refereed to as Paper I]{Ochiai} found through N-body simulations incorporating both planet-planet and planet-star tidal interactions, that the formation rate of binary planets is as much as $\sim 10\%$ of the systems in which orbital crossing among multiple gas giants occurs.  Furthermore, Paper I also predicted the binary orbital separation distribution and the limit of stellarcentric semimajor axis of the binaries beyond which the binary orbits are stable during the main sequence lifetime of solar-type stars.  Thereby, binary planets may have more chance to be detected than exomoons if these theoretical predictions are correct.\footnote{If exomoons can be formed in a different way than \citet{Canup06} considered, exomoons, can be larger and their detection is not so difficult.}

The important point is that the mechanism to form the binary planets proposed by \citet{Ochiai} is one of the natural outcomes of orbital crossing among gas giants.  The gas giants so far discovered in extrasolar planetary systems often have eccentric (say, $e \ga 0.2$) orbits, except for close-in planets where the eccentricities are tidally damped.  It is considered that orbital crossing for the case of a system containing three gas giants or more, is the most likely origin of gas giants in eccentric orbits \citep[e.g.,][]{LI97,MW02,c08,Juric08}.  Typical fates of such three planet systems are ejection of a planet, planet-planet collisions, and planet-star collisions.

However, if planet-star tidal interaction is taken into account, most of the planet-star collisions are replaced by formation of ``hot jupiters."  \citet{nagasawa} and \citet{BN12} found through N-body simulations that hot jupiters are formed in as much as 10-30\% of the systems.  The discovery of retrograde hot jupiters  \citep[e.g.,][]{Narita09,Winn09} strongly suggests that some fraction of hot jupiters were formed through the tidal capture.  

\citet{posi} pointed out that incorporation of planet-planet tidal interactions replaces some of planet-planet collisions with formation of binary planets.  While \citet{posi} assumed arbitrary initial conditions of two giants in closely packed, nearly circular orbits, Paper I considered more appropriate initial conditions consisting of three giants in separated orbits, and found that the formation rate is still as much as $\sim 10\%$.  \citet{Ida13} showed through a planet population synthesis simulation that such orbital crossing among gas giants commonly occurs in systems formed from relatively massive protoplanetary disks.

Possible methods to detect binary gas-giant planets include radial velocity, transit light curves, transit timing variations (TTV), Rossiter-McLaughlin (RM) effect, and gravitational microlensing.  Because  binary planets may be tight binaries, the detection of such binary planets will be through deviation from a single planet fit.  \citet{posi} showed that the radial velocity amplitude of the deviation is too small.  Detection by gravitational microlensing is also possible if a background star passes between the binary planets, but they suggested that the possibility of such events is too low.  We found that TTV signals are also too small.  While detection by RM effect is not ruled out, a bright host star is required, and the number of sufficiently bright stars with transiting gas giants beyond 0.3 AU is limited. Therefore, transit (light curve) observation is the most promising method to detect these binary planets.

In addition, binary planets may be sufficiently numerous to be detected in large transiting planet surveys.  Assuming that the $\sim 10\%$ of planetary systems hosting eccentric gas giants are all a result of orbital instability, and combining this with the predicted $\sim 10\%$ binary planet formation probability, we have that $\sim 1\%$ of systems could host a binary planet.  Taking a typical stellarcentric semi-major axis of 0.5AU, this gives a transit probability of $\sim 1\%$.  Consequently, we expect that approximately 1 in 10000 stars would host a transiting gas giant binary planet pair.  This compares favourably with the $\sim 145000$ stars monitored by Kepler and $\sim 100000$ light curves produced by CoRoT.

Although \citet{posi} also concluded that transit observations are the most promising way to detect binary planets, they did not discuss how planet binarity modifies transit light curves, or calculate detection probabilities.  In addition, they did not calculate long-term tidal evolution of the binaries either. \citet{SA09} calculated the modulation of transit light curves, but as they considered Earth-Moon like systems, focusing on mutual eclipses, their arguments cannot be applied to the binary gas giants that we consider here.

In this paper, based on the detailed theoretical predictions by Paper I, we calculate the effect of planet binarity on transit light curves and discuss the possibility of detection of binary planets. In section \ref{riron}, we summarize the results of Paper I.  In \S \ref{rei}, we predict light curves of possible binary planet systems and discuss the detectability of extrasolar binary planets.  In \S \ref{KepSec}, we apply this work to the problem of detecting binary planets in the CoRoT and Kepler candidates, while \S \ref{kousatsu} is a summary.

\section{THEORETICAL PREDICTIONS}
\label{riron}
Here we briefly summarize the results of theoretical calculations
in Paper I.
Paper I carried out two sets of calculations: 1) N-body simulations
of three gas giants incorporating planet-planet dynamical tides
(as well as planet-star tides) on timescales $\sim 10^7$ years
to investigate tidal capture to form binary planets
and 2) numerical integration for long-term evolution of
the binary orbits due to planet-planet and
planet-star quasi-static tides during main sequence lifetime
of solar type stars ($\sim 10^{10}$ years).

Dynamical behaviors in two planet systems are qualitatively different from 
those with three planets or more.
For the case of two planets, they immediately start orbital crossing
when their initial orbital separation is smaller than a particular critical value,
while close encounters never happen otherwise.
In systems with three planets or more, there is no solid stability boundary.
With modest initial orbital separations, three planet systems can
start orbital crossing after their eccentricities are built up 
over relatively long timescales \citep[e.g.,][]{Chambers}.
Because it is not easy to establish unstable orbital configurations of two planet
systems, Paper I considered the three gas giant planet systems
having equal-mass of $M_{\rm J}$ and radius $2R_{\rm J}$,
where $M_{\rm J}$ and $R_{\rm J}$ are the present values of Jupiter.
The results of the N-body simulations are summarized as follows:
\begin{enumerate}
\item
During close encounters, energy dissipation due to planet-planet tides
often results in formation of a gravitationally bound pair of planets
(binary planets).
Tidal capture usually occurs in the early phase of the orbital crossing
before the planets' stellarcentric eccentricities have not been maximally excited.
The formation rate is $\sim 10\%$ of the runs, almost independent of
initial stellarcentric semimajor axes of the planets.
\item The stellarcentric semimajor axes of the binary barycenters
are comparable to the initial locations.
Their stellarcentric eccentricities are distributed in a broad range
with median values of $\sim 0.15$.
\item The binary planets are tight binaries.
After the orbital circularization and long-term tidal evolution, 
the binary separations
are $\sim 3-5$ times of the sum of the physical radii of the planets ($R_{\rm tot}$).
\end{enumerate}
Hereafter we use the subscripts "0", "1", and "2" to represent the period at
the tidal capture, tidal circularization, and the spin-orbit synchronous state just after quasi-static planet-planet tidal evolution, respectively.
Because tidal interaction is a sensitive function of distance,
the close encounters that lead to tidal capture are usually grazing ones.
So, the binary orbits just after the tidal capture 
have pericenter distance of
$q_{\rm bi,0}\sim R_{\rm tot}-2R_{\rm tot}$ and 
binary orbital eccentricity $e_{\rm bi,0} \sim1$.
The binary separations after tidal circularization are
given by $a_{\rm bi,1} \sim 2q_{\rm bi,0} \sim 2R_{\rm tot}-4R_{\rm tot}$ 
due to conservation of angular momentum
($\sqrt{a_{\rm bi,1}} \sim \sqrt{a_{\rm bi,0}(1-e_{\rm bi,0}^2)}\sim 
\sqrt{2q_{\rm bi,0}}$).

\begin{figure}
\begin{center}
\includegraphics[width=0.9\textwidth]{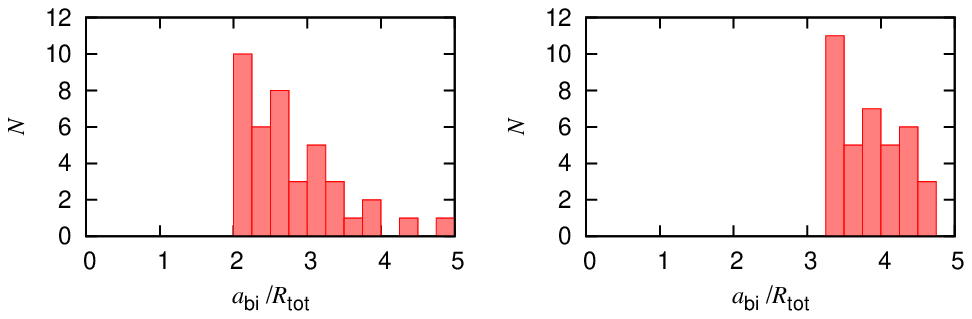}
\end{center}
\vspace{2.5cm}
\caption{
The distributions of binary separations.
The left panel shows the separations ($a_{\rm bi}$)  
just after tidal trapping
followed by binary orbital circularization by 
planet-planet dynamical tides that are
obtained from the N-body simulations conducted in Paper I.
The separations ($a_{\rm bi,1}$) after entering  
the spin-orbit synchronous state as a result of
long-term quasi-static tidal evolution are plotted
in the right panel. 
\label{fig:a_bi}
}
\end{figure}

After the tidal trapping to form binary planets and the binary
orbital circularization due to planet-planet dynamical tides,
the binary separation expands, and the binary planet pair enters a spin-orbit synchronous state 
through quasi-static tidal evolution.
The initial total angular momentum of the circularized binary
is $L_1 = 2 \times (2/5)M_{\rm p} R_{\rm p}^2 \omega_{\rm p}+ \mu a_{\rm bi,1}^2 \Omega_{\rm bi,1}$,
where $\mu = M_{\rm p}/2$, $M_{\rm p}$ is the planetary mass, $R_{\rm p}$ is the planetary radius, $\omega_{\rm p}$ is the spin angular velocity of the
individual planets, $\Omega_{\rm bi,1} = \sqrt{2GM_{\rm p}/a_{\rm bi,1}^3}$,
and $G$ is the gravitational constant.
In the spin-orbit synchronous state after quasi-static planet-planet tidal evolution,
the total angular momentum is
$L_2 = 2 \times (2/5)M_{\rm p} R_{\rm p}^2 \Omega_{\rm bi,2}+ \mu a_{\rm bi,2}^2 \Omega_{\rm bi,2}$.
For initial spin period of 10 hours, $M_{\rm p}=M_{\rm J}$ and $R_{\rm p}=2R_{\rm J}$,
the orbital separation in the spin-orbit synchronous state is given by
\begin{eqnarray}
a_{\rm bi,2} &\sim & a_{\rm bi,1}\left(1+\frac45\sqrt{\frac2{GM_{\rm p}a_{\rm bi,1}}}R_{\rm p}^2\omega_{\rm p}\right)^2 \nonumber \\
&\sim&
a_{\rm bi,1}
\left(1+0.4\bun{a_{\rm bi,1}}{10R_{\rm J}}^{-0.5}\right)^2,
\label{eq:a_bi_2}
\end{eqnarray}
where we assumed $a_{\rm bi,2} \gg R_{\rm p}$.
The distributions of $a_{\rm bi,1}$ and $a_{\rm bi,2}$ obtained by
Paper I are showin in Fig.~\ref{fig:a_bi}.  
At stellarcentric distance $a_{\rm G} \la a_{\rm G, Hill} \simeq 0.2$ AU, 
$a_{\rm bi,2}$ exceeds $r_{\rm H}/3$ where $r_{\rm H}$ is Hill radius ($r_{\rm H} = (M_{\rm p}/3M_\ast)^{1/3}a_{\rm G}$) and the perturbations of the central star
destabilize the binary \citep{Sasaki}. 
At $a_{\rm G} \la a_{\rm G, tide} \simeq 0.4$ AU, planet-star quasi-static tide removes
the binary orbital angular momentum and the binary planets collide with each other
within main sequence phase of solar type stars ($\sim 10^{10}$ years).
Since the planets' gas envelopes fully contract in $10^{8}$ years,
in the quasi-static tidal evolution $R_{\rm p}=R_{\rm J}$ is more appropriate
than $R_{\rm p}=2R_{\rm J}$.
With $R_{\rm p}=R_{\rm J}$, the critical stellarcentric semimajor axis is $a_{\rm G, tide} \sim 0.3$ AU, rather than 0.4 AU.
Therefore, binary planets should be orbitally stable at $a_{\rm G} \ga 0.3$ AU.

Since the timescale to establish the spin-orbit synchronous state
is $\la 10^6$ years and binary separation does not evolve significantly once this state is achieved, except in the case of the orbital destabilizations,
observed binary planets should be in this spin-orbit synchronous state,
that is, binary separations of $3R_{\rm tot}$--$5R_{\rm tot}$, 
which is comparable to the stellar diameter.

In addition, these binary systems should be stable when $a_{\rm G} \ga 0.3$ AU. 
Based on these theoretically predicted orbital parameters,
we discuss the detectability of binary planets by transit observations
in the following sections.

\section{TRANSIT LIGHT CURVES OF BINARY PLANETS}
\label{rei}

We discuss how the transit light curves of binary planets are modified compared with those of a single planet, using the parameters given in the last section, with the aim of providing insight into the types of light curves that can be produced by a binary planet pair.  For simplicity, we take the stellarcentric eccentricity of the binary barycenter as $e_{\rm G}=0$ and the stellarcentric semi-major axis to be $a_{\rm G}=0.4$ AU, which is near the stability limit, $a_{\rm G, tide}$.  Although transit detectability increases with decreasing $a_{\rm G}$, the lifetime of binary planets is shorter.  So, binary planets just outside $a_{\rm G, tide}$ may be the most promising targets.

We show the transit light curves of binary planets 
with equal-mass ($M_{\rm p} = M_{\rm J}$) and equal-radius ($R_{\rm p}$). 
For the radius, we consider two cases: $R_{\rm p} = 2 R_{\rm J}$ and $R_{\rm p} = 1R_{\rm J}$.
We set the orbital separation after the tidal circularization of 
binaries as $a_{\rm bi,1}=2.5 R_{\rm tot}$, 
which is a typical value obtained in Paper I.
From Eq.~(\ref{eq:a_bi_2}),
the orbital separation after the orbit enters the spin-orbit synchronous state
is estimated as $a_{\rm bi,2}\simeq 3.9 R_{\rm tot}$
for $R_{\rm p} = 2 R_{\rm J}$ and $a_{\rm bi,2}\simeq 2.8 R_{\rm tot}$
for $R_{\rm p} = 1 R_{\rm J}$. 
The corresponding binary rotation periods are $\simeq 5.4$ and 3.3 days
for $R_{\rm p} = 2 R_{\rm J}$ and $R_{\rm p} = 1R_{\rm J}$, respectively.
Since $R_{\odot} \sim 10 R_{\rm J}$, the binary separations are
$\sim 0.8$ and $\sim 0.3$ times the stellar diameter for $R_{\rm p} = 2 R_{\rm J}$ 
and $R_{\rm p} = R_{\rm J}$, respectively.
As we pointed out in the last section, these binaries are most likely to be observed in the spin-orbit synchronous state.
So, we assume a spin-orbit synchronous state for the binary planets.
The stellarcentric orbital period is $\sim 92$ days,
assuming a solar-mass central star and $a_{\rm G}=0.4$ AU.
For simplicity, we assume that the stellarcentric and binary orbits are coplanar.

For light curves, limb-darkening is also taken into account, while stellar spots
and pulsation are neglected.  We calculate sample light curves using the method of \citet{Pal2012}, using the quadratic limb darkening model of \citet{Pierce2000}.  For simplicity, we assume the radius, mass and luminosity of the host star are equal to the solar values of $R_\odot$, $L_\odot$, and $M_\odot$.

\begin{figure}
\includegraphics[width=0.9\textwidth]{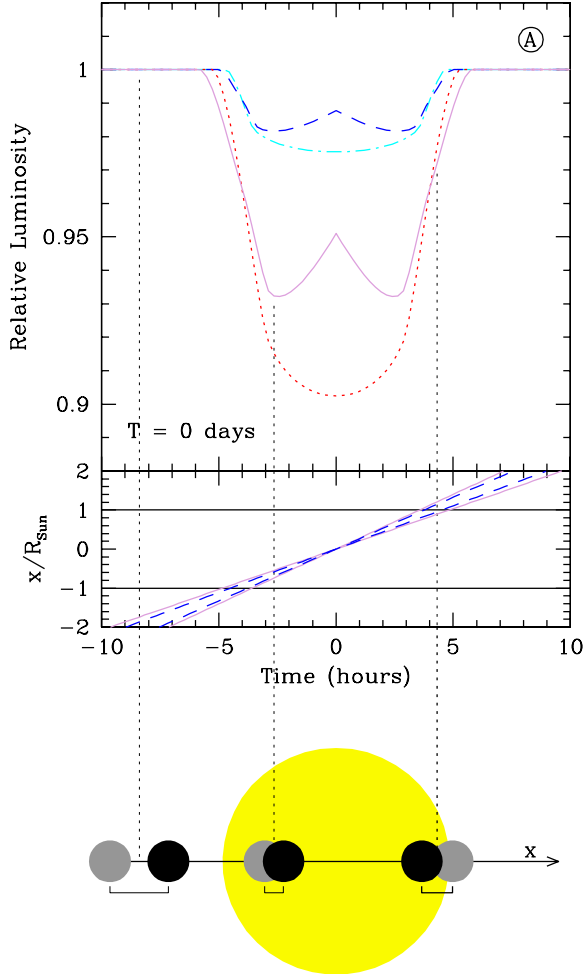}
\vspace{-1.2cm}
\caption{
The top panel shows transit light curves for simulated binary planets transiting a solar-like limb darkened star for the case where the binary's barycenter passes the center of the photosphere as the binary planet undergoes a mutual event along the line-of-sight.  The cases of a $1M_{\rm J}$, $2R_{\rm J}$ binary (solid purple line), $2M_{\rm J}$, $2\sqrt2 R_{\rm J}$ single planet (dotted red line), $1M_{\rm J}$, $1R_{\rm J}$ binary (dashed blue line) and $2M_{\rm J}$, $\sqrt2 R_{\rm J}$ single planet (dash-dotted light-blue line) are shown.  The middle panel gives the positions of the members of the $2R_{\rm J}$ (solid purple line) and $R_{\rm J}$ (dashed blue line) binary pairs along the transit chord as a function of time.  The bottom figure shows the positions of the transiting binary planet (the filled black and grey circles) in front of the photosphere (the filled yellow circle).  The $x$-axis is in the direction of motion of the binary's barycenter and normalized by the Sun's radius, $R_\odot$.  The origin of $x$-axis is taken at the center of photosphere.  The barycenter of binary always remains on the $x$-axis and the two planets move in coplanar orbits.
\label{kurowakusei}
}
\end{figure}

Figure~\ref{kurowakusei} shows an example of a transit light curve along with the positions of the  binary planets, corresponding to the case where the binary's barycenter passes the stellar center.
In this case, the silhouette of the planets overlap (mutual transit) near the transit center.
The top panel shows the transit light curve.
The solid purple line represents the predicted light curve of a binary consisting of planets
with $M_{\rm p}=1M_{\rm J}$ and $R_{\rm p}=2R_{\rm J}$.
For comparison, the light curve for a single planet 
with $M_{\rm p}=2M_{\rm J}$, $R_{\rm p}=2\sqrt2 R_{\rm J}$ at
semimajor axis $a=0.4$ AU is also shown using a dotted red line.  For this case we set $R_{\rm p}=2\sqrt2 R_{\rm J}$, such that
the transit depth is equal to the total transit depth of the two planets.
Similarly, the light curve for a binary with $M_{\rm p}=1M_{\rm J}$ and radius $R_{\rm p}=R_{\rm J}$,
and a single planet with $2M_{\rm J}$ and $\sqrt2 R_{\rm J}$
is also shown using a dashed blue line and a dash-dotted light-blue line respectively.
The middle panel gives the projected positions of the planets.
The $x$-axis is the distance along the transit path
with the origin at the center of the star.
The bottom panel
shows the projected positions of 
the transiting binary planets (the filled black and gray circles)
in the case of $R_{\rm p} = 2R_{\rm J}$
for a number of snapshots during the transit.  
As shown in this illustration, the motion of the two planets around their common barycenter during
the transit is taken into account when calculating the light curve. 

The leading planet of the binary (the filled black circle) reaches the edge of the photosphere at $t\sim -6$ h from the transit center.
Because of limb darkening and the area of the star being blocked by the planet increasing, the light curve gradually goes down with time.
Because there is an off-set due to the binary separation, 
the ingress (the egress) is earlier (later) than for the
single planet fit with $a=0.4$ AU.  For the case where only one transit is observed, and a single planet fit is done for the transit duration, the stellarcentric semimajor axis may be overestimated.  The overestimation depends on the binary orbital phase during ingress and egress.  Although overestimation is only slight in the case of Fig.~\ref{kurowakusei}, the maximum overestimation of the orbital period is a factor of 2, and thus, the semimajor axis would be overestimated by a factor of $2^{2/3} \sim 1.6$. For the case where multiple transits are observed, the orbital period is known, and thus the semi-major axis can be accurately determined.
  
\begin{figure}
\includegraphics[width=\textwidth]{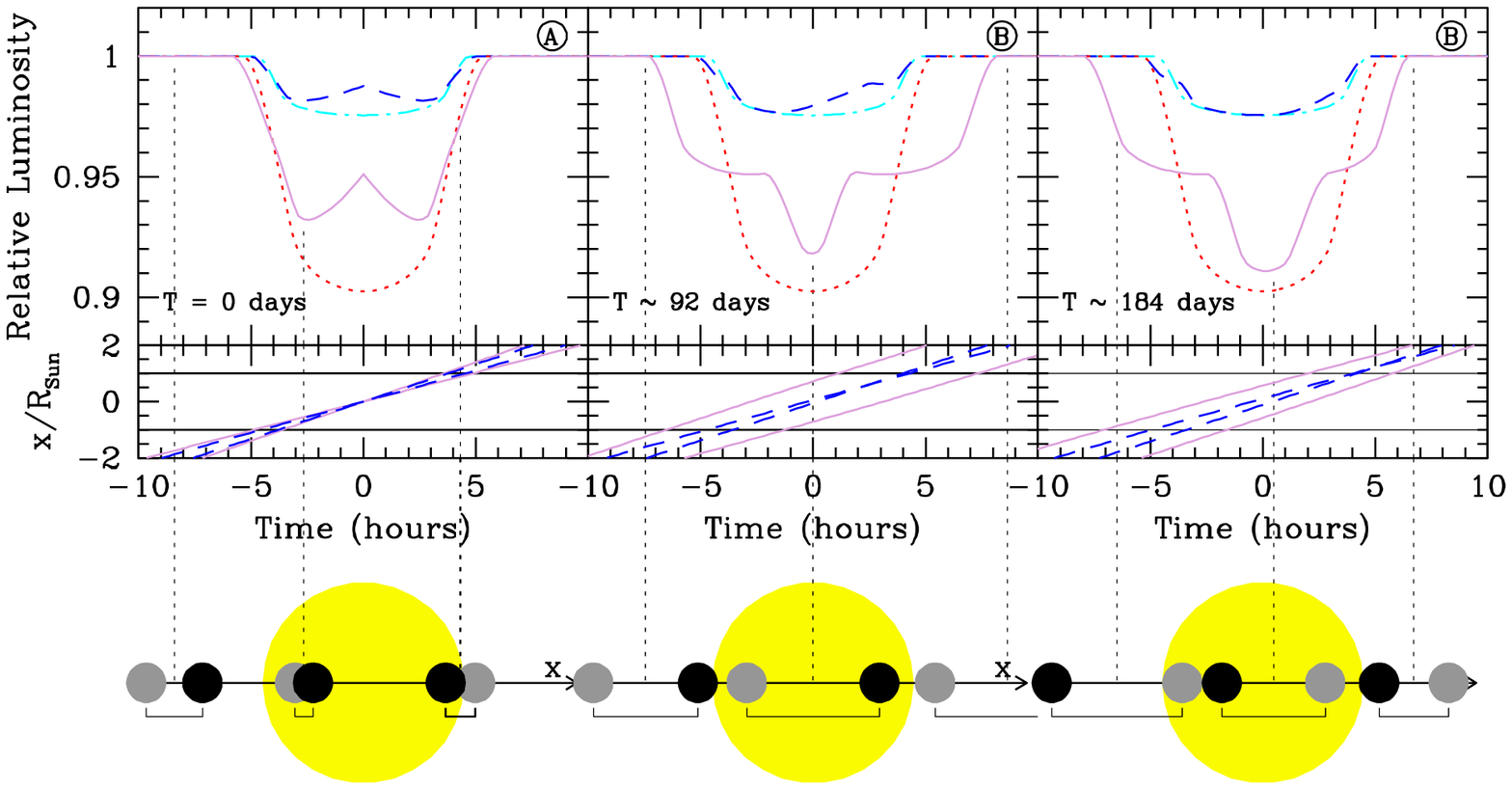}\vspace{-8cm}\\
\includegraphics[width=\textwidth]{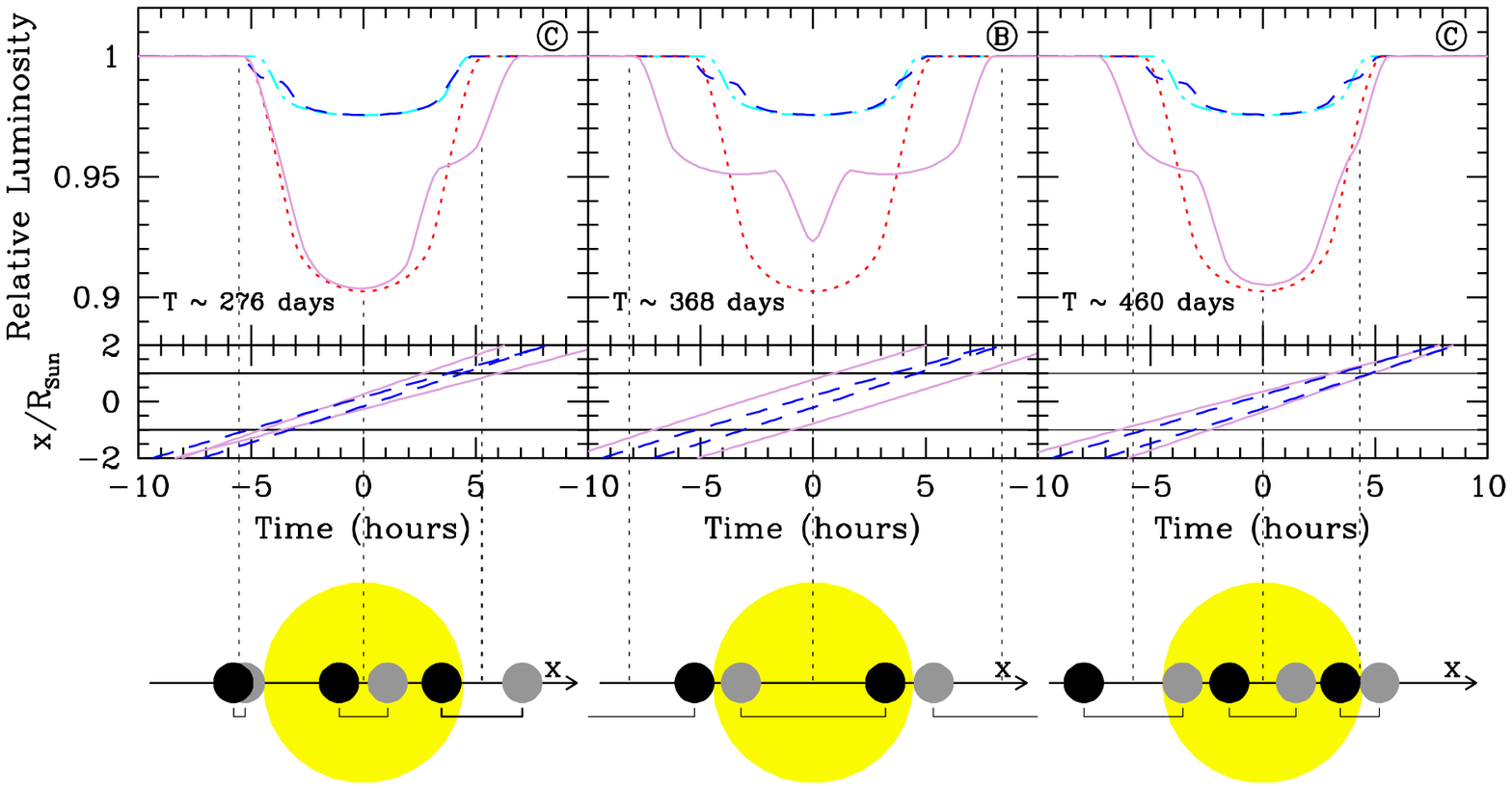}
\vspace{-8cm}
\caption{
A sequence of six consecutive transit light curves
and the projected positions of the transiting binary planets,
for the case of Fig.~\ref{kurowakusei}.
The upper-left panel is the same as Fig.~\ref{kurowakusei}.
We set $t=0$ when the binary's barycenter passes the stellar surface 
center in all figures.
The light curve in the upper-left panel is case A,
the light curves in the upper-right and upper/lower-middle panels
are case B, and
those in the lower-right/left panels are case C.\label{sougoshoku04}
}
\end{figure}

At the transit center ($t\sim0$), 
the light curve of the binary planet pair shows a bump due to a
mutual eclipse between the planets.
Depending on the positions of the two planets in their mutual orbit and orbital separation,
a dip often appears at the transit center,
instead of the bump (Figure~\ref{sougoshoku04}).  
Such a bump or a dip never occurs for a single planet transit,
except for a bump resulting from a planet passing in front of a stellar spot.

The most pronounced property of the light curves of
binary planets is variation in the light curves from transit-to-transit.
Figure~\ref{sougoshoku04} shows the light curves
for a sequence of six consecutive transits
of the same binary system as that shown in Fig.~\ref{kurowakusei}.
As can be seen, each light curve is different.
The upper-left panel is the same as Figure~\ref{kurowakusei},
which we call ``case A". 

The three panels (the upper-right and upper/lower-middle panels) show a dip.
Because the orbital separation is comparable to the stellar
diameter for $R_{\rm p} = 2R_{\rm J}$, the duration in which
only one planet of the binary is transiting and that in which
both planets are transiting are often comparable.
In that case, the transit light curves show a deep dip (transit
by both planets) near the transit center,
sandwiched by relatively shallow transit
(transit by one planet).
We call this ``case B".
In this case, the two-planet transit occurs without a
mutual transit.
However, since one or the other of the planets is transiting a highly limb darkened region,
the transit depth is slightly shallower than for the single planet case.

Also in the lower-right/left panels, the case of the transit of one planet 
and that of two planets can coexist.  
However, since the projected binary separations are smaller 
than those in case B, the curves show a ``step" rather than a ``dip."
Since the two planet transit  
is not affected by the limb darkening, 
the maximum transit depth is similar to the single planet fit.
We call this case ``case C".

\begin{figure}
\includegraphics[width=0.9\textwidth]{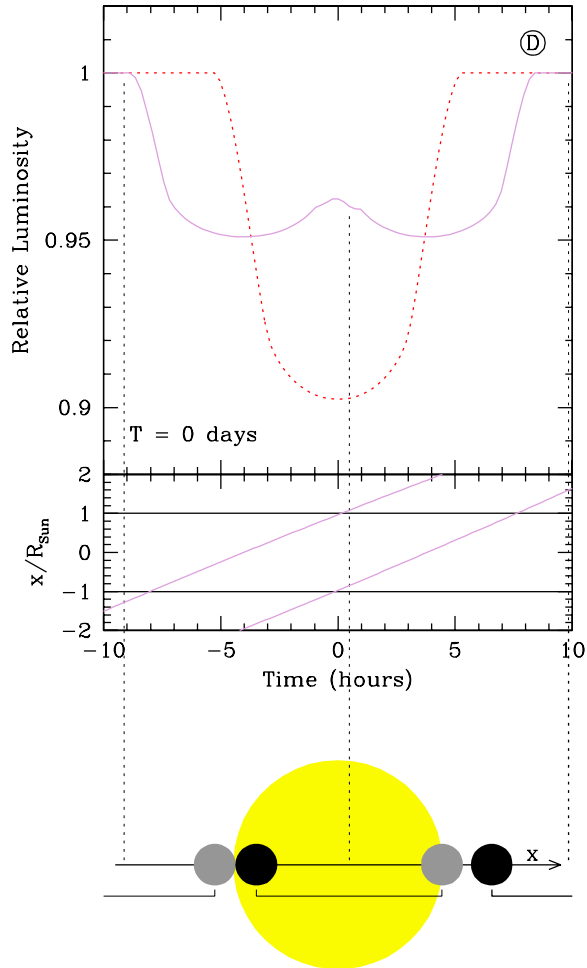}
\vspace{-1.2cm}
\caption{
An example of a transit light curve of a binary planet in case D.
We set $R_{\rm p}=2R_{\rm J}$ and the orbital separation that 
is 1.25 times larger than that in Figure~\ref{sougoshoku04} ($\simeq 3.9R_{\rm tot}$).
\label{smallbump}
}
\end{figure}

Note that if the orbital separation is slightly larger, 
such that one planet enters the transit just at the time 
when the other leaves it, 
the transit curve has a bump at the transit center with a shallower 
transit depth than case A (Figure~\ref{smallbump}).
We call this case ``case D".
We summarize cases A to D in table \ref{ABCDmatome}.

\begin{figure}
\includegraphics[width=\textwidth]{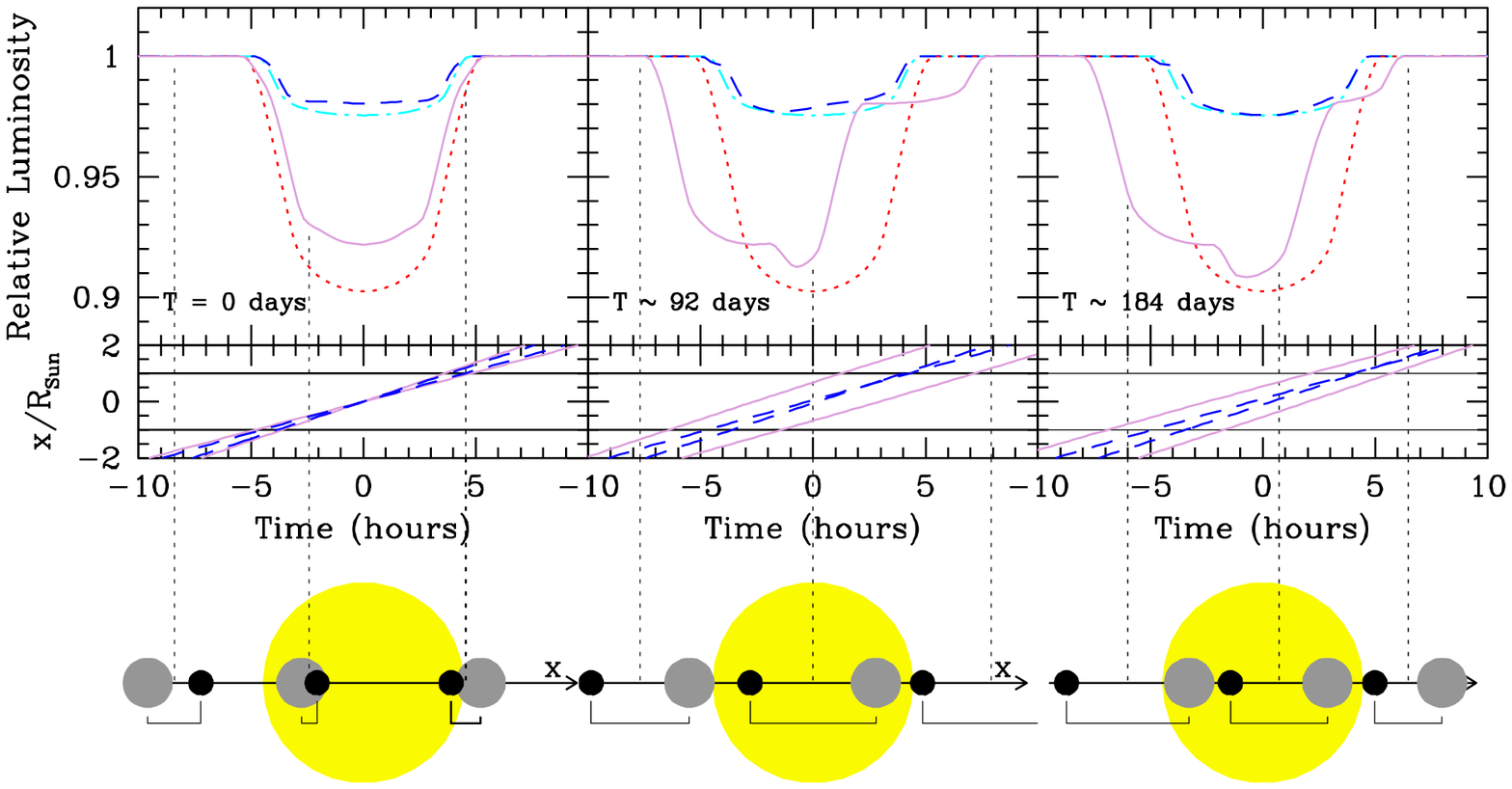}\vspace{-8cm}\\
\includegraphics[width=\textwidth]{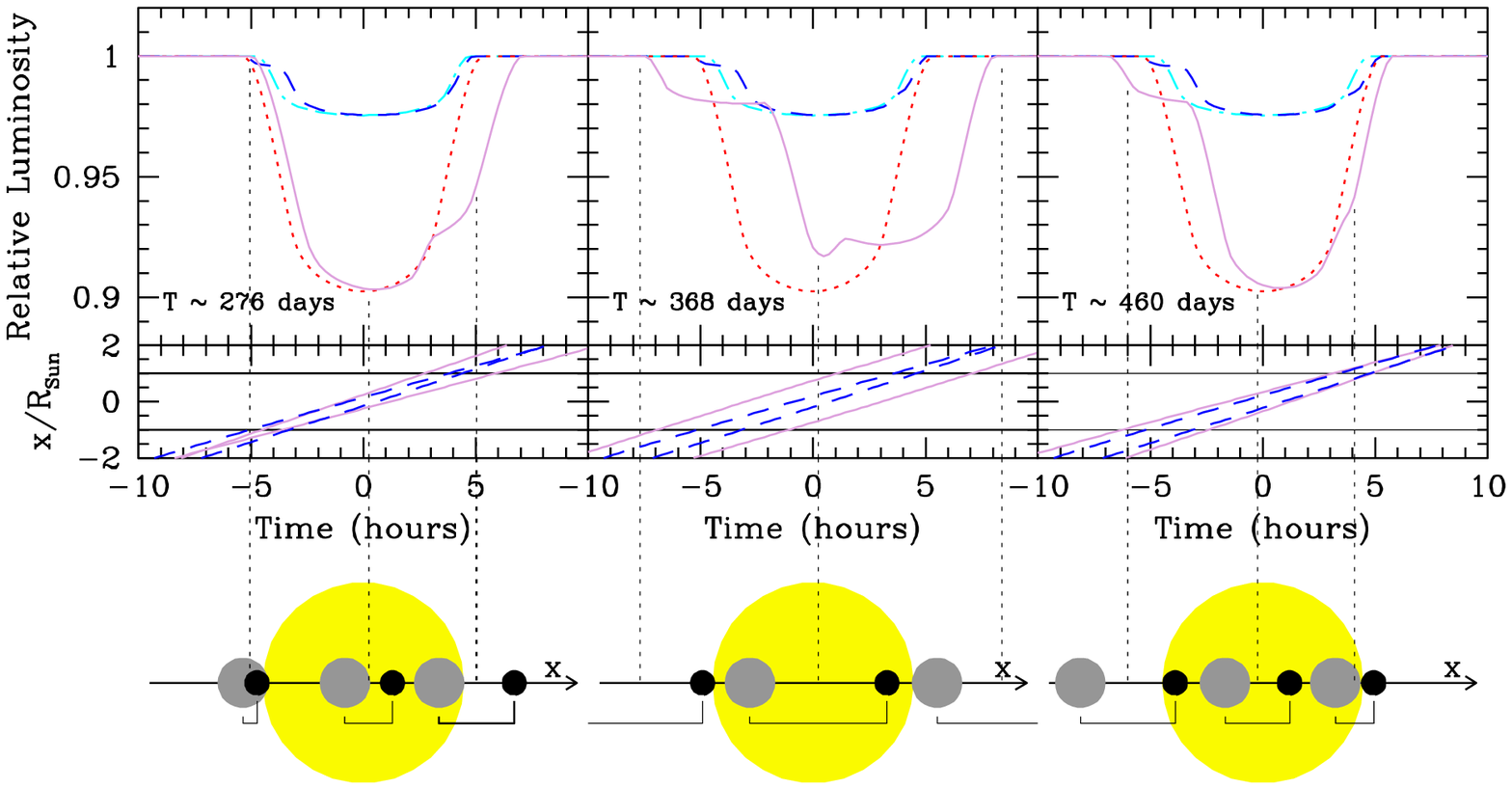}
\vspace{-8cm}
\caption{
Same as Figure~\ref{sougoshoku04},
except that the ratio of the physical radii of the binary planets is 2:1,
keeping the total cross-section the same.
\label{sougoshoku04r21}
}
\end{figure}

Note that if we consider
the binary systems with non-equal $R_{\rm p}$, the modulation of
light curves becomes less pronounced. 
Figure~\ref{sougoshoku04r21} is 
the same as Figure~\ref{sougoshoku04},
except that the ratio of the physical radii of the binary planets is 2:1,
keeping the total cross-section the same.
In this non-equal $R_{\rm p}$ case, statistical techniques for detecting exomoons
may become necessary.

Case A is characterized by a mutual transit.  From \citet{SA09}, 
the detection probability of binary planets with a mutual transit is
\begin{eqnarray}
p=p_1p_2,
\end{eqnarray}
where $p_1=t_{\rm{obs}}/T_{\rm K}$ ($T_{\rm K}$
is a stellarcentric Keplerian period of the binary barycenter) is the probability
for the binary center to pass across the surface of the central star
during the observational duration $t_{\rm obs}$, 
$p_2=t_{\rm{E}}/T_{\rm{bi}}$ is the fraction of transits for which a mutual transit occurs during the transit duration $t_{\rm{E}}=2R_\odot/v_{\rm{K}}$ ($p_2=1$ for $T_{\rm{bi}}<t_{\rm{E}}$), 
where $T_{\rm{bi}}$ is the Keplerian period of the binary system
and $v_{\rm{K}}$ is the stellarcentric Keplerian velocity of the binary barycenter.
For $a_{\rm{bi}}=2.5(R_i+R_j)=10R_{\rm J}$,
$p \sim 0.1$.
So, the probability of case A is low.
Note that \citet{Hirano} detected a rare mutual transit,
although it is for a pair of non-bounded planets.

If the total mass of the binary system is determined
through RV measurement,
their mean bulk density can be calculated.
However, as stated in the above,
a single planet fit based on the transit duration
leads to fitting parameters of $M_{\rm p}=2M_{\rm J}$ and $R_{\rm p} = \sqrt{2}R_{\rm J}$.
As a result the calculated bulk density can be up to $(\sqrt{2})^3/2=\sqrt{2}$ times 
smaller than the real one, with the exact factor depending on the derived impact parameter for both fits (see section \ref{detect_corot} for an example).  So, binary planets may be mis-classified as
inflated exoplanets, if a single planet fit is applied.
The tidal stability limit $a_{\rm G, tide}$
evaluated by the overestimated $R_{\rm p}$ is artificially large, so if inflated exoplanets are located inside of $a_{\rm G, tide}$
(but outside of $a_{\rm G, Hill}$), it is worth considering the possibility that the object is really a binary planet pair. 

The changes in the light curves in a sequence of consecutive transits
may be the most pronounced signal of planet binarity.
However, if the stellarcentric orbital period is close to
an integer number of times of the binary period and the total number of
observed transit is relatively small, the transit-to-transit changes may not be significant.
So, the detection of perturbations to the light curve shape for each individual transit is also important.

To see if such changes could be detectable, we have conducted a preliminary search of the open access CoRoT and Kepler transiting planet data.  This investigation is discussed in the next section.

\section{DETECTABILITY OF BINARY PLANETS IN ARCHIVED COROT AND KEPLER DATA} 
\label{KepSec}

To investigate if binary planets analogous to those investigated in this paper could be detected using current technology, we focussed on data from the CoRoT and Kepler satellites.  To provide context, both these missions, and the types of planets they were designed to detect will be discussed in turn.

CoRoT is an ESA-led mission with the aim of using a 27cm diameter space telescope to detect transiting planets larger than Earth, in short period orbits as well as monitoring and characterising stars \citep{Auvergne2009}.  As the CoRoT satellite was in orbit of the Earth it suffered from thermal effects related to its orbital phase which led to predictable data errors.  In addition, this satellite monitored a range of fields, located in the galactic centre and anti-center, with one observing run per field.  Before the mission ended, nine long runs (150-90 days) and three short runs (20 days) towards fields in the galactic centre were completed along with six long runs, five short runs, and one intermediate length run (50 days) towards the galactic anti-center.  As a result of this mission, $\sim$ 100,000 light curves have been released and 27 planets have been detected.  Given that less than a year's worth of contiguous data was available for each  candidate, it is unsurprising that most planets detected by CoRoT had small semi-major axes and short periods.

On the other hand, Kepler is a Nasa-led mission with the aim of using a 0.95m diameter telescope to discover an Earth-twin around a Sun-like star \citep{Borucki2008}.  To ensure less noise and a longer time baseline, Kepler was placed in an Earth-trailing orbit and monitored $\sim$ 145,000
stars in one particular field.  Apart from small gaps in the data due to satellite rotations, data downlinks and problems such as safe modes and coronal mass ejections, the data is continuous.  As a result of the high data quality and the long time baseline, multiple transits for each candidate were routinely collected, allowing planetary orbital period to be derived and allowing sensitivity to sub-earth sized planets.  These factors resulted in a much larger number of Kepler planet candidates (4234) as well as confirmed planets (978).\footnote{Retrieved from http://kepler.nasa.gov/, 4$^{th}$ August, 2014.}

In this context, detection of binary planets using CoRoT and Kepler data will be discussed in turn.  In particular, we show that simple light curve fitting programs can successfully identify binary planet candidates using a possible binary planet candidate from the CoRoT data set and confirmed binary star pair, transiting a brighter star, from the Kepler data set.  Then, as the Kepler data set is a much richer place to search for binary planets, we conduct a simulation to demonstrate that Kepler quality data is sufficient to detect these planets. 

\subsection{DETECTABILITY OF BINARY PLANETS IN COROT DATA} 
\label{detect_corot}

\begin{figure}[tb]
\vspace{-7cm}
\includegraphics[width=0.9\textwidth]{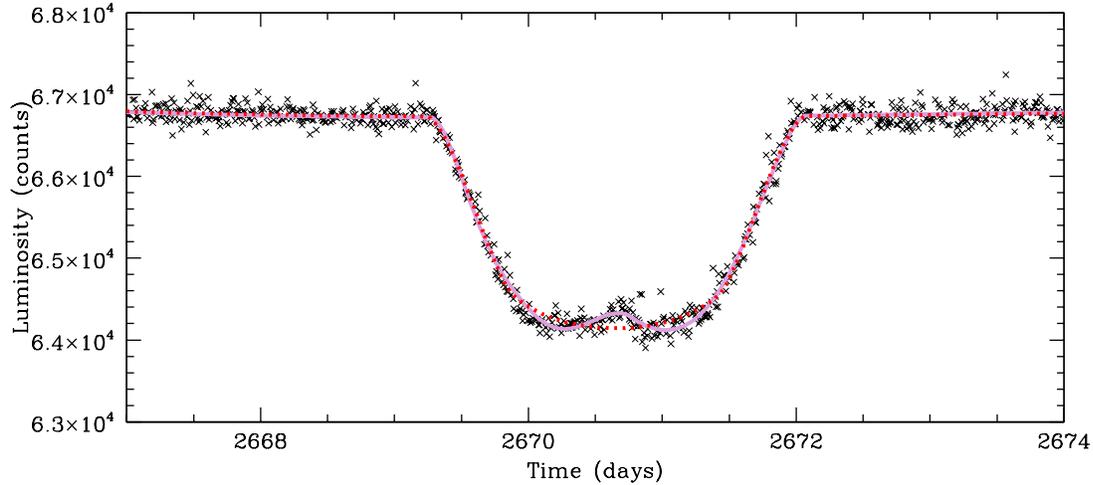}
\caption{Light curve for CoRoT target SRc01 E2 1066, where only the data points (black crosses) with status flag set to 0 are plotted.  The single planet (dotted red) and binary planet fit (solid purple) are also shown.  The best fit values corresponding to these fits are given in table~\ref{CoRoTFitTable}.}
\label{CoRoTFitFigure}
\end{figure}

To investigate binary planet detection, the CoRoT candidates were checked by eye.  One interesting object that we discovered is CoRoT SRc01 E2 1066, which is described in detail in \citet{erikson}.  The transit has a relative depth of 4\% and duration of 66 hours (see figure~\ref{CoRoTFitFigure}).  From the long transit duration, \citet{erikson} suggested that this event might be a transit of an evolved or dwarf star by a distant gas giant planet, where, by chance, the planet occulted a stellar active region (spot) at the centre of the transit.  But this event could also be due to a transit of a binary gas giant planet pair with smaller stellarcentric semi-major axis.  

To check this claim we performed a single planet (dotted red) and binary planet (solid purple) fit to this data (see figure~\ref{CoRoTFitFigure}).  The single planet and binary planet models were calculated using the light curve simulation code of \citet{Pal2012}, where, to account for long term trends in the light curve, the out-of-transit light curve was modelled using a cubic.  For simplicity, we assume that the binary orbit normal is circular and is perpendicular to the line-of-sight and the chord made by the planet-planet barycenter across the star.  The best fit models were then determined by locating the minimum residual using a simplex minimisation routine, with the fit improvement, calculated using the change in Bayesian Information Criterion (BIC), defined as
\begin{equation}
\Delta BIC = n \ln\left(\frac{RSS_{sing}}{n}\right) - \ln\left(\frac{RSS_{bin}}{n}\right) - k \ln (n)
\end{equation}
where $RSS_{sing}$ and $RSS_{bin}$ are the residual sum of squares for the single planet and binary planet cases respectively, $k$ is the number of extra degrees of freedom, five for this work, and $n$ the number of data points used for the fit, in this case 1333. The derived parameters for the single and binary planet cases are shown in table~\ref{CoRoTFitTable}, noting that the BIC value improves by $\sim80$ when the binary, as opposed to the single planet model, is used, where a value of 10 is robust.  As can be seen from table~\ref{CoRoTFitTable}, the binary planet fit yields orbital and physical parameters for the putative binary tantalisingly similar to those derived in paper I.  However, as only one transit is available (as this was a short run) it is not possible to rule out the starspot case.  If more data were available for this target, its true nature could be determined by e.g. testing for long term trends due to light curve modulation due to starspots, resulting from stellar rotation.  In addition, if another transit were detected, it would help differentiate between these cases by first, placing a strong constraint on the orbital period (which is different in the binary and non-binary planet cases) and second indicating if there were transit-to-transit variations in transit width, depth and shape (see figures~\ref{sougoshoku04} to \ref{sougoshoku04r21}) as is likely for the binary planet case.  These issues are discussed in greater detail in section~\ref{det_starspots}.

\subsection{DETECTABILITY OF BINARY PLANETS IN KEPLER DATA} 

While no visually obvious binary planets are present in the Kepler candidate and binary star data sets, a transiting binary star pair was discovered \citep{Carter2011}.  In this system, the central star has radius $2.0254 \pm 0.0098$ $R_\odot$ while the transiting pair have radius $0.2543 \pm 0.0014$ $R_\odot$ and $0.2318 \pm 0.0013$ $R_\odot$ respectively, and the binary has a semi-major axis of 4.729 $R_\odot$.  As the host star is between 3000 and 5000 times brighter than the members of the transiting binary pair, and the radius ratios are similar to those of the planetary binaries described in this paper and in paper I, these light curves should be approximately analogous to those of the planetary case, and we can test our fitting code on a real transiting binary system.  

\begin{figure}[tb]
\includegraphics[width=0.9\textwidth]{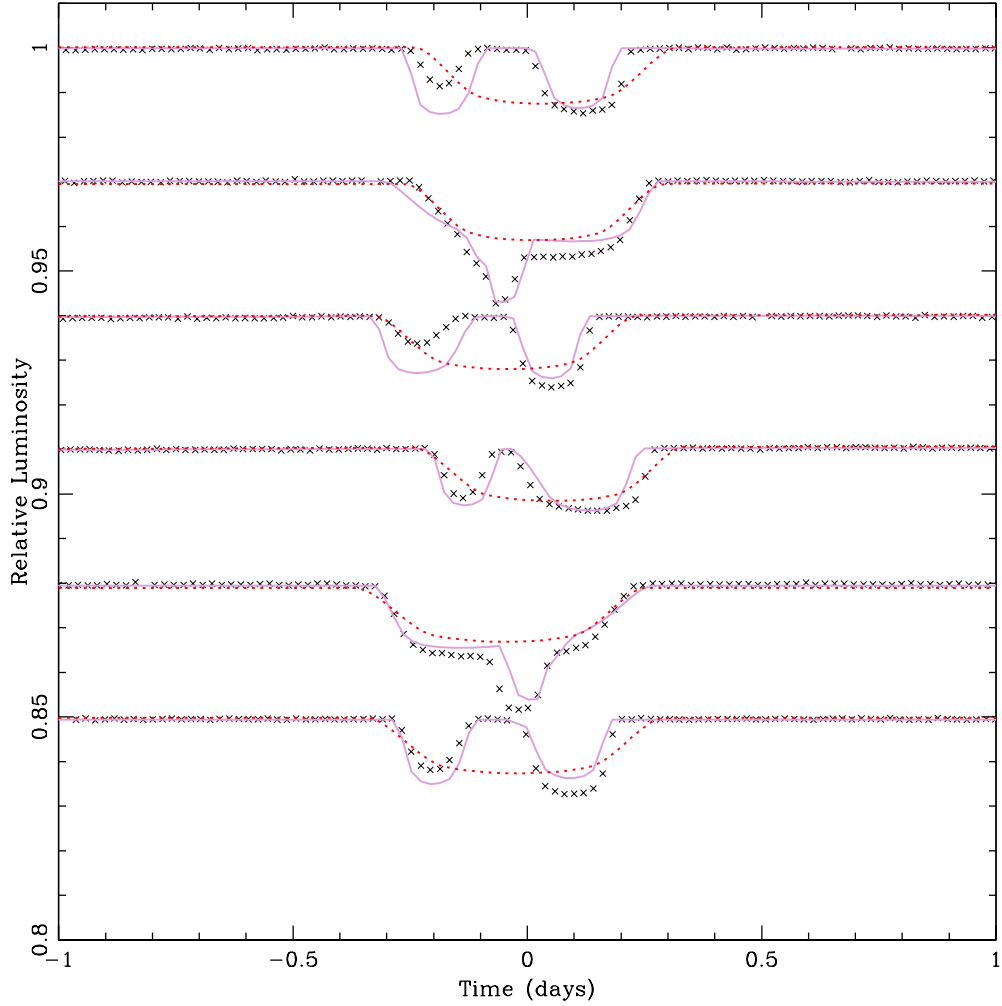}
\caption{The first six transits of the binary star pair in the KOI-126 system (black crosses) along with the best fit single star (dotted red line) and binary star (purple line) model calculated using our simplified fitting code.}
\label{CarterTransitSequenceFigure}
\end{figure}

Example transits from this system are shown in figure~\ref{CarterTransitSequenceFigure}.  Note how they look markedly different from those from a single object transit and similar to some of the sample transits shown in figures~\ref{sougoshoku04} and \ref{smallbump}.  We fit this data with our simplified fitting model, given by the solid (purple) line.  As this transiting binary system has an inclined orbit, and the impact parameter for the transit is large, we do not expect the light curve and the model to exactly match.  In particular, this binary orbit is inclined such that the leading star transits closer to the stellar limb than the trailing star and is the reason why the transit duration and depth observed for the leading star is shorter and shallower than predicted by the fit and the transit duration and depth observed for the trailing star is longer and deeper than predicted by the fit.  While the fit isn't perfect, the binary system is unequivocally detected with a reduction in the BIC of over 2000, which indicates that our simple model still can successfully detect binary systems for inclined binaries and high transit impact parameter systems.


To investigate if binary planets analogous to those investigated in this paper could be detected in real Kepler observations, a preliminary investigation was conducted. Focussing on the long cadence data, we selected a Kepler candidate, KOI 3681.01 (KIC 2581316), showing an anomalously large radius (22 Earth radii), transit duration (21.3 hr) and with a host magnitude close to the 12$^{\textrm{th}}$ magnitude Kepler target specifications (11.69), with the aim of demonstrating that it is possible to robustly detect the difference between a binary planet transit light curve and a transit light curve due to an inflated single planet. To do this, we simulated realistic noisy binary planet transit light curves using two different noise sources, KIC 2581316/KOI 3681, a 11.69 magnitude star and KIC 9517393/KOI 2076, a 15.3 magnitude host star, and then  tried to detect the planet binary by fitting a single planet and binary planet light curve model.  The method used to produce these light curves is described in the next section.


\subsubsection{Simulating Noisy Binary Planet Lightcurves} 

Simulated realistic binary planet light curves were constructed from real Kepler light curves in five stages.  First, long cadence data was downloaded from the NASA Exoplanet Archive\footnote{http://exoplanetarchive.ipac.caltech.edu/index.html} corresponding to quarters 0 to 16.  Second, all planetary transits were completely removed.  Third a sequence of binary planet transits was simulated.  Fourth, out-of-transit data was used to give the simulated transits realistic noise.  Finally, the simulations and the original data were stitched together, where a cubic was added to the simulation to ensure that the gradient and value of the endpoints matched.  The final four stages will be discussed in detail.

For both KIC 2581316 and KIC 9517393, all planetary transits were first identified.  To ensure that no signal corresponding to real moons of planetary candidates in these systems remained, estimates of the planetary masses were used to calculate reasonable values for the Hill sphere, and all data corresponding to the transit of the planet or the Hill sphere was removed.

\begin{figure}[tb]
\includegraphics[width=0.9\textwidth]{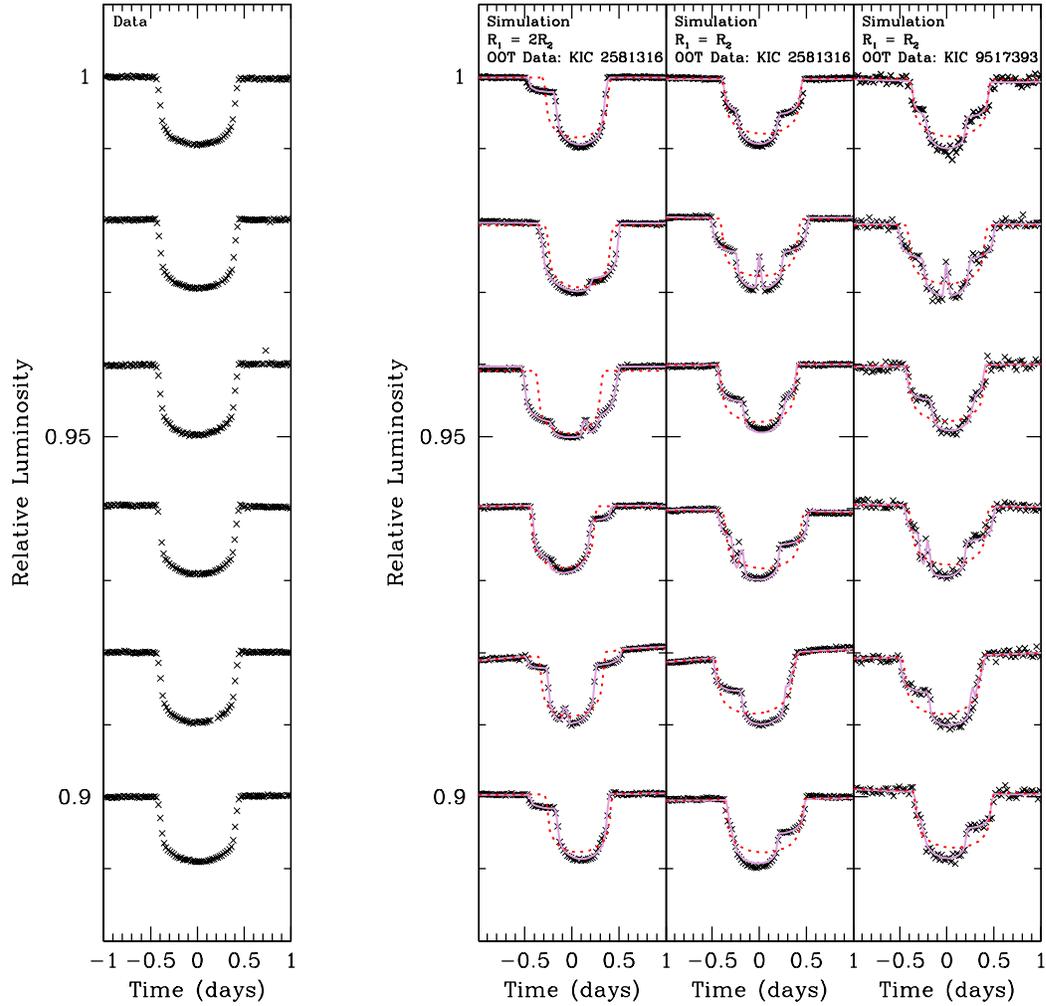}
\caption{The real transits of KOI 3681.01 (left) from quarters 2, 5, 7, 9, 12 and 14, along with simulated transits for the case of a binary pair with radius ratio 2:1, an equal radius binary pair, and an equal radius binary pair where out of transit data from KIC 9517393 was used to contaminate the light curve (right).  The single planet (dotted red) and binary planet (solid purple) fits are shown for the simulated data.}
\label{KeplerTransitSequenceFigure}
\end{figure}

Then, using the code of \citet{Pal2012}, we simulated sequences of transits of binary planet pairs.  Following Paper I, we simulated two classes of binary planet pairs, one where both components had equal radii and one where the radii of one of the components was twice that of the second.  To ensure that the transit depth for the binary case was similar to that observed for the real candidate (see figure~\ref{KeplerTransitSequenceFigure}), for the equal radius case we set $R_1 = R_2 = 0.0629R_*$ while for the other case we set $R_1=0.0795R_*$ and $R_2=0.0398R_*$.  Also, the semi-major axis of the planet binary was taken to be $R_*$ corresponding to a typical binary planet separation (see section~\ref{rei}), and the transit velocity was altered to approximately match the transit duration.  In addition, following the analysis in section~\ref{rei}, the planets are assumed to be equal mass and, for simplicity, these systems are assumed to have zero eccentricity, be coplanar with the planet orbit and have impact parameter equal to that of KOI 3681.01.

To ensure these light curves displayed realistic photometric noise, we randomly selected sections of out of transit light curve from Q1-15 data from KIC 2581316 for the case of low noise and KIC 9517393 for the case of high noise.  Our simulated light curves were then multiplied by these sections of of data to produce noisy binary planet light curves.

Finally, the simulated light curves were stitched into the original data.  To ensure that the light curves were continuous, a cubic was added to each transit light curve such that the gradient and value at both edges of the simulated transit light curve matched the gradient and value at the edges of the original data.

Using this method we simulated 50 six transit\footnote{KOI 3681.01 shows seven transits while KOI 2076.02 shows six.} sequences of light curves for these systems (see figure~\ref{KeplerTransitSequenceFigure} for examples), where the initial true anomaly was randomised for each sequence.



\subsubsection{Determining if the Binary Planets were Detectable}

%

\begin{figure}[tb]
\includegraphics[width=0.9\textwidth]{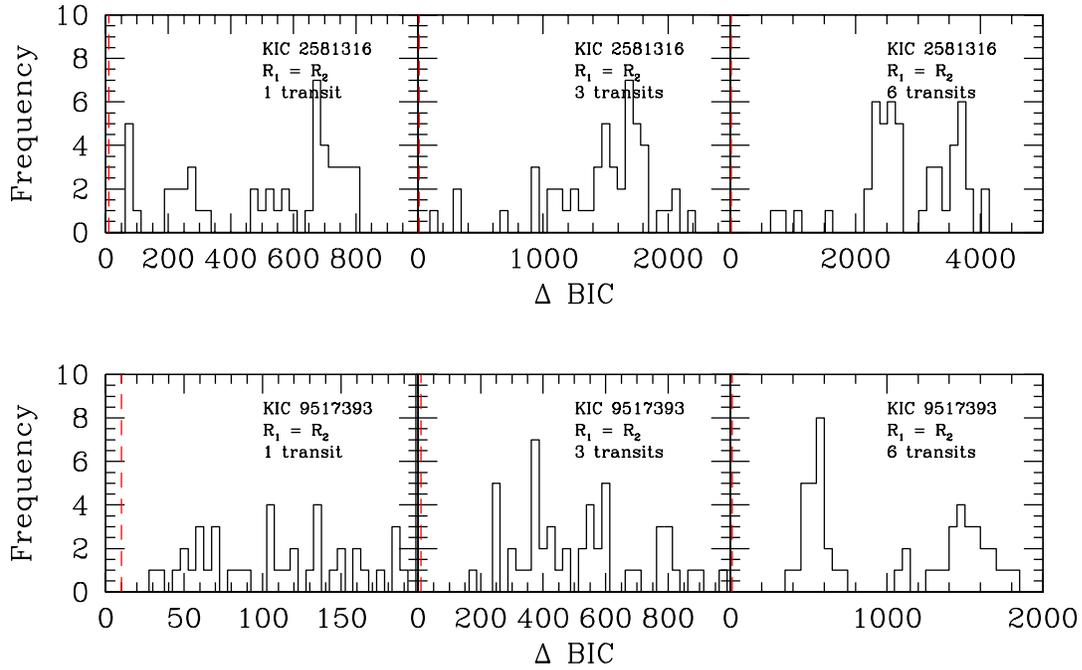}
\caption{Distribution of the decrease in \textbf{BIC} for the set of simulations of an equal radius binary planet along with the robust detection limit} (red dashed line).
\label{KeplerEqualResultsFigure}
\end{figure}

\begin{figure}[tb]
\includegraphics[width=0.9\textwidth]{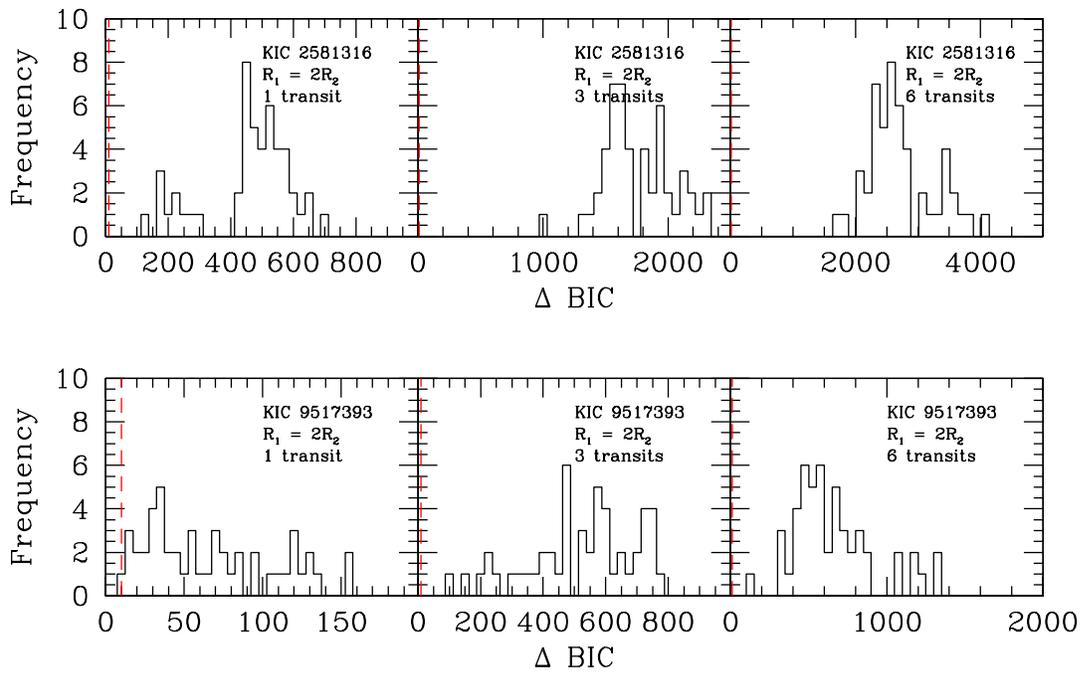}
\caption{Distribution of the decrease BIC for the set of simulations of a binary planet with radius ratio 2:1, again plotted with the robust detection limit (red dashed line).}
\label{KeplerUnequalResultsFigure}
\end{figure}

To determine if planet binarity was detectable in these simulated light curves, using our light curve fitting code, we fitted each sequence of transits with a single planet and binary planet model and recorded the difference in the BIC.  To determine the effect of data length we repeated the process including only one and only three transits.  Some example fits are shown in figure~\ref{KeplerTransitSequenceFigure} and the results are shown in figures~\ref{KeplerEqualResultsFigure} and \ref{KeplerUnequalResultsFigure}.  All binary planets were detected, nearly all, robustly.

\begin{figure}[tb]
\includegraphics[width=0.9\textwidth]{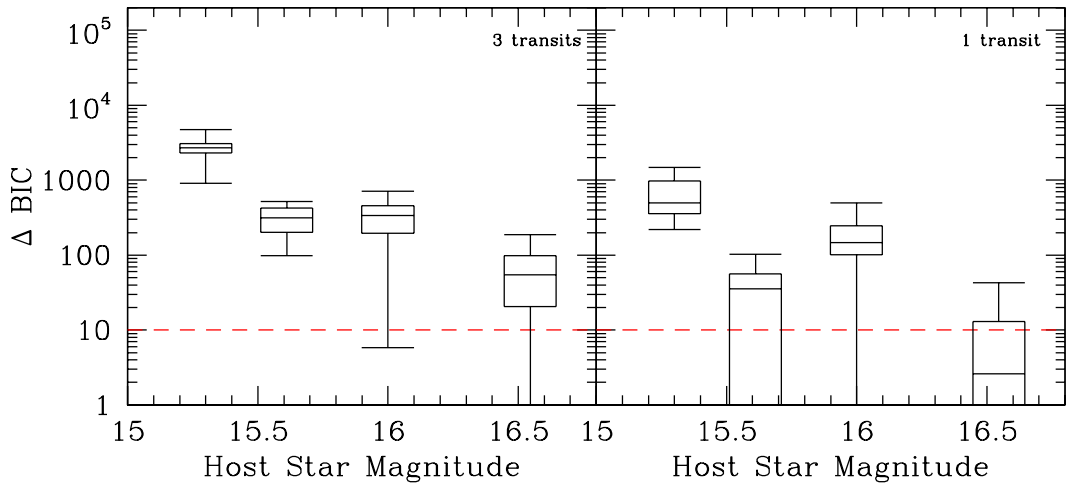}
\caption{Box plot showing the minimum, first quartile, median, third quartile and maximum decrease in BIC for the set of simulations of an equal radius binary planet for a range of different stellar magnitudes along with the robust detection limit (red dashed line).}
\label{KeplerMagnitudesFigure}
\end{figure}

To investigate the behaviour of the detection threshold as a function of host star magnitude, three additional stars were chosen from the Kepler catalog, KIC 12121701 (magnitude 15.611), KIC 8827930 (magnitude 15.999) and KIC 2438406 (magnitude 16.546).  While these results apply to these specific systems, the trend should be indicative of the true trend.  The analysis described previously was repeated for the equal radius ratio and one or three transit cases, with the results plotted in figure~\ref{KeplerMagnitudesFigure} along with the results for KIC 9517393.  As can be seen, in our sample, binary planets are robustly detected for stars with magnitude less than 16 and 15.5 for the three and one transit cases. 

\begin{figure}[tb]
\includegraphics[width=0.9\textwidth]{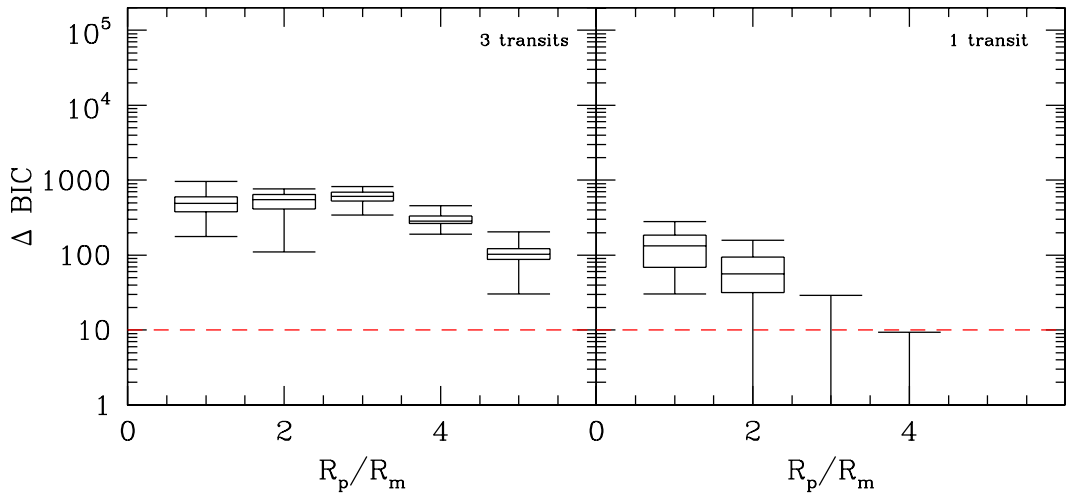}
\caption{Box plot showing the minimum, first quartile, median, third quartile and maximum decrease in BIC for the set of simulations of binary planets with a range of radius ratios with noise taken from KIC 9517393 with the robust detection limit (red dashed line).}
\label{KeplerRadiiFigure}
\end{figure}

In addition, binary planets may exist with radius ratios outside of those investigated in Paper I. To investigate the detectability of such systems, simulations with a range of radius ratios were constructed using photometric noise from KIC9517393, the 15.3 magnitude target, and analysed using our code.  As can be seen from figure~\ref{KeplerRadiiFigure}, for the case where one transit is observed, binary planets are robustly detected for radius ratios smaller than 2:1 while for the case of three observed transits, even planet pairs with radius ratio 5:1 are robustly detected.

From these simulation results we show that a range of binary planets, analogous to those simulated in this paper could be practically detected in Kepler data even for the case where the star is dim (15$^{th}$ magnitude) and the number of transits is small.  In particular we show that binary planets are more detectable around host stars with low relative photometric noise compared to high photometric noise  targets.  In addition, we show that detectability improves when the number of observed transits increases.

Finally while this analysis does include many important physical factors e.g. realistic photometric noise, transit to transit variation, it does not include the effect of spot crossing events in the light curve.  As discussed in section~\ref{rei}, possible spot crossing events may closely mimic the effect of gas giant binary planet transit, so this topic warrants discussion.

\subsection{THE EFFECT OF STARSPOTS ON BINARY PLANET DETECTION} 
\label{det_starspots}

One factor which may hamper the detection of binary planets is the presence of spot crossing events in a transit light curve.  We propose that it should be possible to differentiate between a single planet which orbits a spotty star and a binary planet given sufficient numbers of transits by either confirming the presence of spots or by providing supporting evidence for a binary planet.  In particular this can be done in  a number of ways including:
\begin{enumerate}
\item \emph{Investigate if the timing of spot crossing events corresponds to a physically realistic star:}  For the case where the planetary orbit is inclined with respect to the stellar spin axis, the transit chord may cross one or more active latitudes.  As a result, spot crossing events are more likely to appear on the same part of different transits light curves, with the position corresponding to the location of the active latitudes \citep[e.g.][]{SanchisOjeda2011}.  In addition, it has been suggested that interactions between the planet and the star can also lead to  correlation between the position of the planet and the position of active regions.  This phenomenon may be present and detectable in some systems \citep[e.g.][]{Pagano2009,Herrero2013} but not others \citep[e.g.][]{Miller2012,Scandariato2013}, however, as the planets of interest for this work are distant (outside 0.3AU), such an interaction is likely to be very weak or absent. Consequently, the presence of light curve perturbations that cannot be predicted, but occur with higher probability in certain sections of the light curve, could mean strong evidence for starspots being the cause.   However for the case of binary planet pairs, the size and location of bumps in the light curve should relate to the sizes of the planets and will always correspond to a physically realistic planet model.  
\item  \emph{Use trends in the out-of-transit light curve to confirm presence of spots:}  Starspots also cause changes to the out-of-transit light curve.  \citet{SanchisOjeda2012} suggested corellating long term light curve modulations and spot crossing events to determine relative inclination in multi planet systems, but the same processes could be used to provide evidence for or against a spot being the cause of a particular light curve feature.
\item   \emph{Image the stellar surface to confirm presence of spots:}  For a possible evolved host star, they are known to host large starspots.  Such spots have been imaged \citep[e.g.,][]{Vogt83,Vogt87,Strassmeier02}.  For some host stars this may be an option to determine the presence of spots.
\item  \emph{Compare measured planetary mass to predicted mass:}  As mentioned previously, for the single planet case, the mass derived from radial velocity is likely to match the measured radii, while for binary planets that have been incorrectly classified, it will be systematically low.
\end{enumerate}

Similar to the simulations presented, these arguments indicate that the detectability of binary planet pairs increases as the number of transits increases.  In addition, observations that are likely to be taken to try to prove planetary nature e.g. RV observations can also be used to determine likelihood of a given candidate being a binary planet.

\section{CONCLUSIONS}
\label{kousatsu}

In this paper, we have discussed the possibility of the observational detection of extrasolar binary planets (gravitationally bound pairs
of gas giant planets) by transit observations,
based on the results of N-body simulations on the tidal capture between
gas giants and 
calculations of long-term tidal evolution after the capture performed by Paper I.
Paper I showed that the formation probability of a planetary binary is as much as $\sim 10\%$
almost independent of stellarcentric semimajor axis, of the binary ($a_{\rm G}$) 
and predicted that the typical binary separation is 3--5 times the sum of 
physical radii of the planets and the binary planets are tidally stable for $\sim 10^{10}$ years
if $a_{\rm G} \ga 0.3$ AU. 

Using these constraints, we have modelled transit light curves of physically plausible binary planets.
These light curves have a deep dip, a big bump or a step or two separate dips, and are noticeably and statistically different 
from those of a single planet.  Furthermore, the transit shape changes from transit to transit, compared to the single planet case.  Because of these features, the transits of binary planets might be classified as
false positives.\footnote{
Light curves that change irregularly  have been already discovered \citep[e.g.,][]{Barnes}, although this object is inside of the tidal stability limit and 
would not be a binary planet.}
If RV measurement is also available,
the bulk density can be estimated.
The single planet fit for two equi-sized binary planets can give a lower bulk density
than the real value.
Thus true binary planets could also have been classified as ``inflated" planets
if  $a_{\rm G} \ga 0.3$ AU.

We show that the CoRoT target SRc01 E2 1066 is well fit by a binary planet model and put forward the alternate scenario that it could also be due to the transit of a binary planet in addition to the starspot scenario proposed by \citet{erikson}.   In addition, we show that binary planets may be present in, and would be detectable in the Kepler data set and are most detectable where the host star shows little noise and a number of transits are available.  In addition, for host stars with magnitude less that 15, we show that a broad range of binary planets are robustly detectable, even for the case of one or a few observed transits. Prompted by this preliminary analysis we propose that we should do an accurate reanalysis of the irregular changing light curves of Kepler and CoRoT planets and planet candidates orbiting beyond 0.4AU from their central star, where over 100 candidates and false positives exist.

\acknowledgements

We are grateful for the offer of the gravitational lensing calculation program by Takahiro Sumi.
We also thank Tristan Guillot and Rosemary Mardling for discussions on observations of binary planets.  In addition, we thank Jessie Christiansen for advice on the Kepler pipeline.  We would also like to thank an anonymous referee for helpful comments which improved the quality of this paper.  This research was supported by a grant for JSPS (23103005) Grant-in-aid for Scientific Research on Innovative Areas.  KML was supported by JSPS KAKENHI Grant Number: 24-02764.

\clearpage

\clearpage

\begin{table}
\begin{tabular}{ccccc}
\tableline
 case & transit depth & transit duration & transit shape \\
\tableline\tableline
 A     & half          &  similar           &  bump due to mutual transit  \\
 B      & slightly shallower  & longer & deep dip \\
 C      & similar      & longer & asymmetric step \\
 D      & half          & twice & bump due to wide separation\\
 \tableline
 \end{tabular}
\caption{The light curve properties of a binary planet pair compared with those of an analogous single planet with the same stellarcentric semi-major axis and total cross-sectional area.
\label{ABCDmatome}
}
 \end{table}
 
\begin{table}[tb]
\begin{tabular}{lc}
\hline
\multicolumn{2}{c}{Single planet fit}\\
\hline
Impact parameter & 0.573043 \\
Transit velocity ($R_*$/day) &  0.738503 \\
Semi-major axis (AU) & 25 \\
Planet radius ($R_*$) &  0.187059 \\
\hline
\multicolumn{2}{c}{Binary planet fit}\\
\hline
Impact parameter & 0.847337 \\
Transit velocity ($R_*$/day) &  1.044480 \\
Semi-major axis (AU) & 12.5 \\
Planet 1 radius ($R_*$) & 0.22087 \\
Planet 2 radius ($R_*$) & 0.156546 \\
Binary semi-major axis ($R_*$) & 0.984286 \\
Mid-transit true anomaly & 1.216958 \\
\hline
\end{tabular}
\caption{The best fit values corresponding to the single planet and binary planet fits shown in figure~\ref{CoRoTFitFigure}.  The semi-major axis listed is calculated assuming a Sun-like host star and a circular orbit of the planet or planet binary around the star.  If, as suggested by \citet{erikson}, the host star is evolved, the semi-major axes calculated for the fits will be much lower.}
\label{CoRoTFitTable}
\end{table}

\end{document}